\documentclass[11pt]{article}
\pdfoutput=1 
\usepackage{jheppub}

\usepackage{color}
\usepackage{amsmath}

\usepackage{bbm}
\usepackage{verbatim}   
\usepackage{subfigure}  
\usepackage{acronym}

\usepackage{amsfonts}
\usepackage{amssymb}
\usepackage{mathrsfs}
\usepackage{graphicx}
\usepackage{multirow}
 \usepackage{slashed}

 {

\newcommand{\der}{\mathrm{d}}

\def\bar#1{\overline{#1}}

\def\inv{^{\raise.15ex\hbox{${\scriptscriptstyle -}$}\kern-.05em 1}}
  
\def\lbar{{\lower.35ex\hbox{$\mathchar'26$}\mkern-10mu\lambda}}

\let\<=\langle
\let\>=\rangle

\let\+=\uparrow

\def\OO{\mathcal{O}}

\title{\huge Towards Cogenesis via Asymmetric Freeze-in:\\[5pt] 
\LARGE The $\chi$ Who Came-in from the Cold}

\author{James Unwin}

\affiliation{Department of Physics, University of Notre Dame, Notre Dame, IN 46556}
\affiliation{Kavli Institute for Theoretical Physics, University of California, Santa Barbara, CA 93106}
\abstract{
In models of freeze-in the dark matter (DM) is decoupled from the visible sector and initially has a depleted number density. The hidden and visible sectors are connected only via a feeble portal interaction by which DM can be produced. Asymmetric freeze-in (AFI) combines this scenario with ideas from asymmetric DM and provides a potential cogenesis mechanism. However, it has been argued that existing AFI models do not produce suitably large particle asymmetries due to cancellations which arises because the mediator state remains in thermal equilibrium. We examine AFI via an out-of-equilibrium mediator and using a simple scalar model show that in this case sizeable asymmetries may be generated. 
}

\setcounter{tocdepth}{1}

\begin{document}

\hfill \vspace{-5mm}  NSF-KITP-14-055

\maketitle

\vspace{5mm} 

\section{Introduction}

Asymmetric dark matter (ADM), see e.g.~\cite{ADM,Hall:2010jx,Cui:2011ab,symC}, provides an interesting alternative to the traditional WIMP scenario. In these models the DM carries a conserved quantum number $X$ and its relic density is determined by an asymmetry between the DM and its antiparticle in direct analogy to the mechanism which sets the present day density of baryons. Moreover, if one supposes that the DM and visible sector are connected through portal operators which violate $B$, $L$ and $X$ but conserve some linear combination, e.g.~$B-L+X$, then the observed coincidence $\Omega_{\rm{DM}}\approx5\Omega_{B}$ may be explained by linking the asymmetries in the hidden and visible sectors, provided that $m_{\rm{DM}}\sim m_{\rm proton}$. This is in stark contrast to conventional DM scenarios, in which the relic abundances of DM and baryons are set by distinct mechanisms. The focus of this paper is a distinct class of ADM models: Asymmetric Freeze-in (AFI) \cite{Hall:2010jx}.

In models of DM freeze-in \cite{Hall,FI} it is generally assumed that the hidden and visible sectors are both separately in thermal equilibrium at different temperatures, with the hidden sector being cooler. Correspondingly, the DM states $\chi$ initially have a depleted number density relative to the visible sector states. The temperature difference between the sectors drives intersector energy exchange via some portal operator and the number density of $\chi$ moves towards equilibrium, or `freezes-in'. In AFI it is proposed that particle asymmetries may be generated during the freeze-in production of DM. The dominant production of $\chi$ typically occurs once the temperature drops below the mass of the mediator state. The portal operator must be sufficiently feeble so that the sectors do not equilibrate in order for the freeze-in production to set the relic density, otherwise it will simply be determined by usual freeze-out. For freeze-in via two-body decays and scattering the coupling should be $\lesssim10^{-7}$ \cite{Hall:2010jx}.

 If the $\chi$ states carry an (approximately) conserved quantum number, analogous to baryon number, and the processes which produce $\chi$ and $\overline{\chi}$ feature a CP-violating phase, then an asymmetry can, in principle, be generated in both the hidden and visible sectors (cogenesis) during freeze-in \cite{Hall:2010jx}. In order to satisfy the requirement that the relic density is set primarily by the asymmetry it is essential that the $\chi\overline{\chi}$ pairs efficiently annihilate leaving only the small residual amount of $\chi$ states due to the particle asymmetry, see e.g.~\cite{symC}.  Annihilation of the symmetric component directly to the visible sector requires relatively large intersector couplings which would lead to sector equilibration. Therefore the removal of the symmetric component of the DM typically requires additional light states in the hidden sector into which the DM can annihilate, and relatively strong interactions between states in the hidden sector.

Whilst it might be expected that simple models of AFI can lead to successful cogenesis, a full analysis of the Boltzmann equations reveals that unforeseen cancellations occur in the minimal model \cite{Hook:2011tk} which results in the asymmetry generation being greatly suppressed. Subsequently, it was suggested that if the mediator involved in the CP-violating process is in thermal equilibrium then generally asymmetries can not be generated \cite{Garbrecht:2013iga}. Drawing on this result, here we consider the scalar AFI model studied in \cite{Hall:2010jx,Hook:2011tk} and examine the case in which the mediator state is out-of-equilibrium. We restrict our attention to this simple scalar model for clarity and to allow for easy comparison; it is expected that this model should exhibit the main features of the AFI mechanism. It should be straightforward to replace the scalar bath states with Standard Model fermions, as this generally leads to only small deviations in the Boltzmann equations which can be typically neglected. A comprehensive analysis, including thermal effects and fermions, is left for future work.

The paper is structured as follows:   In Sect.~\ref{S2} we study explicitly the Boltzmann equations for the scalar AFI model examined in \cite{Hall:2010jx,Hook:2011tk}, but where the mediator is out-of-equilibrium with the other states. Subsequently, in Sect.~\ref{S3}, we derive the Boltzmann equation for the asymmetry and calculate the asymmetric yield. Further, we highlight how the efficiency of the freeze-in mechanism depends on the temperature difference between sectors and argue that washout effects can be negligible for this class of models. The main result we derive is the asymmetric yield for the case that the visible sector, mediator, and DM each have different temperatures. In Sect.~\ref{S4} we analyse the form of the asymmetric yield and argue that if the mediator is out-of-equilibrium, then sizeable asymmetries of comparable magnitude to the baryon asymmetry can be generated. Further, we show that the asymmetry vanishes at leading order if the mediator is brought into equilibrium with the visible sector, in agreement with \cite{Hook:2011tk,Garbrecht:2013iga}. A summary of results and potential directions for further work are given in Sect.~\ref{S5}.


\section{Boltzmann equations for Asymmetric Freeze-in}
\label{S2}

Suppose that the visible sector thermal bath at temperature $T$ is composed of the following states $B_b,~B_2,~B_3$. We take $B_2$ and $B_3$ to be real scalar fields, whilst $B_b$ is a complex scalar which carries a conserved quantum number.  The state $B_b$ is to be identified with a visible sector state carrying $L$ or $B$ (e.g.~a lepton or quark).  In addition, we introduce a complex scalar $\chi$, also carrying the conserved quantum number, which plays the role of the DM, and a mediator state $\phi$ which is a real scalar. The $\chi$ ($\phi$) are in thermal equilibrium at temperature $T_\chi$ ($T_\phi$), but are out-of-equilibrium with the visible sector thermal bath $T_\phi,T_\chi\neq T$, and connected to the visible sector only via feeble interactions. The relevant aspects of the model we shall consider are described by the following Lagrangian (details of the construction are given in Appendix \ref{A1}) 
\begin{equation}
\begin{aligned}
\mathcal{L}=\mu\Big(
\lambda \phi B_b^* \chi
+\lambda' \phi B_2B_3\Big)
+\lambda'' B_2B_3 B_b^* \chi
+{\rm h.c.}\cdots~,
\label{L}
\end{aligned}
\end{equation}
where we omit terms which will not affect the Boltzmann equations. The full Lagrangian has a global U(1) symmetry associated with the conserved charge of $B_b$ and $\chi$.

\subsection{Boltzmann equation for $\chi$}

The Boltzmann equations provide a semi-classical approach for modelling the evolution of number densities of particle species in cosmology and have been used extensively in the study of baryogenesis and leptogenesis (see e.g.~\cite{Kolb:1979qa,Edsjo:1997bg,Kolb:1990vq}). In this section we study explicitly the Boltzmann equations which describe the change in the number densities $n_{\chi}$, $n_{\overline{\chi}}$ of the DM states $\chi$, $\bar \chi$, with the aim of depicting the evolution of the asymmetry $n_{\chi}-n_{\overline{\chi}}~$  in Sect.~\ref{S3}.
The Boltzmann equation for $\chi$ is given by
\begin{equation}
\begin{aligned}
\dot{n}_\chi+3H n_\chi=~&
\Lambda^{\phi}_{b\chi}
\Big[
|M|^2_{\phi\rightarrow b \chi}f_\phi -|M|^2_{b \chi \rightarrow \phi}f_bf_\chi
\Big]\\
&+
\Lambda^{23}_{b\chi}
\Big[
|M|^2_{2 3\rightarrow b \chi}f_2f_3 -|M|^2_{ b \chi \rightarrow 2 3}f_b f_\chi
\Big]\\
&+
\Lambda^{\overline{b\chi}}_{b\chi}
\Big[
|M|^2_{\overline{b\chi}\rightarrow b \chi}f_{\overline{b}}f_{\overline{\chi}}
 -|M|^2_{ b \chi \rightarrow \overline{b\chi}}f_bf_\chi
\Big]
-~\mathcal{R}_\chi,
\end{aligned}
\label{a}
\end{equation}
where we adopt the compact notation
\begin{equation}
\begin{aligned}
\Lambda^{\alpha\beta\cdots}_{ij\cdots} &=\int \der \Pi_\alpha \der \Pi_\beta \cdots \der \Pi_i  \der \Pi_j  \cdots (2\pi)^4\delta^{(4)}(p_\alpha+p_\beta+\cdots-p_i-p_j-\cdots),
\end{aligned}
\label{E1}
\end{equation}
and for conciseness we refer to the bath states by their subscripts. The first line of eq.~(\ref{a}) accounts for $\chi$ number changing decays and the back-reaction, while the terms which follow describe $2\rightarrow2$ processes. The inclusion of on-shell $2\rightarrow2$ processes results in a double-counting problem, a standard solution to which is to perform a real intermediate state (RIS) subtraction \cite{Kolb:1979qa}. The RIS subtraction is accounted for in the term $\mathcal{R}_\chi$ and is given by:

\begin{equation}
\begin{aligned}
 \mathcal{R}_\chi  &=\Lambda^{23}_{b\chi}
\Big[
|M|^2_{2 3\rightarrow b \chi} \Big|_{\rm{RIS}}f_2f_3 -|M|^2_{ b \chi \rightarrow 2 3} \Big|_{\rm{RIS}}f_b f_\chi
\Big]\\
&~+
\Lambda^{\overline{b\chi}}_{b\chi}
\Big[
|M|^2_{\overline{b\chi}\rightarrow b \chi} \Big|_{\rm{RIS}}f_{\overline{b}}f_{\overline{\chi}}
 -|M|^2_{ b \chi \rightarrow \overline{b\chi}} \Big|_{\rm{RIS}}f_bf_\chi
\Big],
\end{aligned}
\label{RIS}
\end{equation} 
where the subscript RIS indicates that the propagator is on-shell. Note that the right-side of eq.~(\ref{a}) also gives $\dot n_b+3Hn_b$,  the evolution of the $b$ number density.

As the states are in sector-wise equilibrium they are described, neglecting statistical factors, by Maxwell-Boltzmann distributions 
\begin{equation}
f_2\simeq\exp\left(-\frac{E_2}{T}\right),
\qquad
f_3\simeq\exp\left(-\frac{E_3}{T}\right),
\qquad
f_\phi\simeq\exp\left(-\frac{E_\phi}{T_\phi}\right).
\end{equation} 
The complex scalars may have chemical potentials \(\mu_\phi,~\mu_\chi\),  and thus their distributions are of the form
\begin{equation}
f_b\simeq\exp\left(-\frac{E_b-\mu_b}{T}\right),
\hspace{2cm}
f_\chi\simeq\exp\left(-\frac{E_\chi-\mu_\chi}{T_\chi}\right).
\end{equation} 
The distributions for the corresponding antiparticle states are identical except with opposing chemical potentials ($\mu\rightarrow-\mu$).

\subsection{Real intermediate state subtraction}

First we consider the interplay between the decays and the RIS subtraction, leaving aside the scattering terms, and we denote this subset of terms $\mathcal{D}_{\chi}$.  From eq.~(\ref{a}) \& (\ref{RIS}) we have
\begin{equation}
\begin{aligned}
\mathcal{D}_{\chi}\equiv&~\Lambda^{\phi}_{b\chi}
\Big[
|M|^2_{\phi\rightarrow b \chi}f_\phi -|M|^2_{b \chi \rightarrow \phi}f_bf_\chi
\Big]\\
&~-\Lambda^{23}_{b\chi}
\Big[
|M|^2_{ 23\rightarrow  b\chi}  \Big|_{\rm{RIS}} f_2f_3 -|M|^2_{ b\chi \rightarrow  23} \Big|_{\rm{RIS}} f_b f_\chi 
\Big]\\
&~-
\Lambda^{\overline{b\chi}}_{b\chi}
\Big[
|M|^2_{ \overline{b\chi} \rightarrow b\chi}  \Big|_{\rm{RIS}} f_{\overline{b}}f_{\overline{\chi}}
 -|M|^2_{ b\chi \rightarrow  \overline{b\chi}}  \Big|_{\rm{RIS}} f_bf_\chi
\Big].
\end{aligned}
\end{equation}
The most elegant way to proceed is to use the following relation (derived in Appendix \ref{A2}) 
\begin{equation}
\Lambda^{ij}_{b\chi}\frac{|M|^2_{ij\rightarrow \alpha\beta}  \Big|_{\rm{RIS}}}{|M|^2_{\phi\rightarrow \alpha\beta}}=\Lambda^{\phi}_{\alpha\beta}\frac{\Gamma_{\phi\overline{ij}}}{\Gamma},
\label{2.6}
\end{equation}
from which it follows that
\begin{equation}
\begin{aligned}
\mathcal{D}_{\chi}=&~\Lambda^{\phi}_{b\chi}
\Big[
|M|^2_{\phi\rightarrow b \chi}f_\phi -|M|^2_{b \chi \rightarrow \phi}f_bf_\chi
\Big]\\
&-\Lambda^{\phi}_{b\chi} \frac{\Gamma_{\phi23}}{\Gamma} 
\Big[|M|^2_{ \phi \rightarrow b\chi}
f_2f_3 - |M|^2_{ b\chi \rightarrow \phi} f_b f_\chi
\Big]\\
&-\Lambda^{\phi}_{b\chi}
\Big[ |M|^2_{ \phi \rightarrow b\chi}  \frac{\Gamma_{\phi b\chi}}{\Gamma} f_{\overline{b}}f_{\overline{\chi}}- |M|^2_{ b\chi \rightarrow \phi}  \frac{\Gamma_{\phi\overline{b\chi}}}{\Gamma} f_bf_\chi
\Big].
\label{2.9}
\end{aligned}
\end{equation}
\begin{figure}[t!]
\begin{center}
\includegraphics[height=30mm]{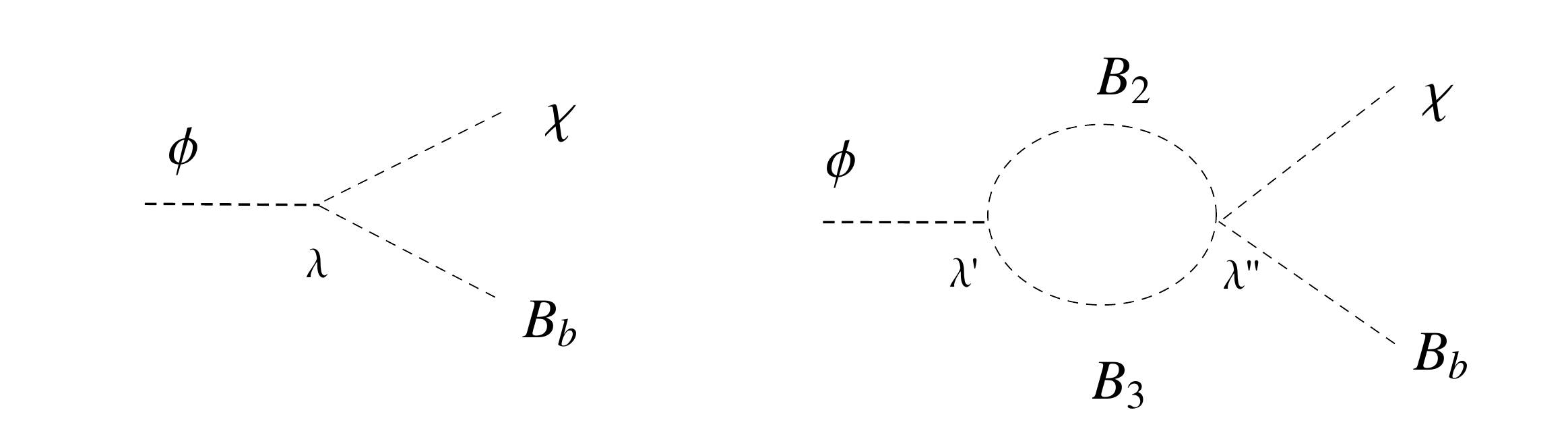}
\vspace{-5mm}
\caption{Interference between the above diagrams gives rise to CP-violation.} 
\label{Fig2}
\end{center}
\end{figure}
Interference between diagrams which contribute towards the process $\phi\rightarrow b\chi$ (and $\phi\rightarrow \overline{b\chi}$), as shown in Fig.~\ref{Fig2}, leads to CP-violation. This can be parameterised as follows
\begin{equation}
\begin{aligned}
|M|^{2}_{\phi \rightarrow b\chi} &=|M|^{2}_{\overline{b\chi} \rightarrow \phi} =\left(\frac{1+\epsilon}{2}\right)(\mu\lambda)^2,
\hspace{1cm}
|M|^{2}_{\phi \rightarrow \overline{b\chi}} &=|M|^{2}_{b\chi\rightarrow  \phi}=\left(\frac{1-\epsilon}{2}\right)(\mu\lambda)^2,
\end{aligned}
\end{equation}
where $\epsilon$ quantifies the CP asymmetry of the interactions. The equality between matrix elements is due to unitarity and CPT (see e.g.~\cite{Kolb:1979qa,Kolb:1990vq}). The matrix elements involving only bath states are of the form $|M|^{2}_{\phi \leftrightarrow 23}=(\mu\lambda')^2$. The form of the CP-violation is given by
\begin{equation}
\begin{aligned}
\epsilon&\equiv~\frac{\Gamma_{\phi b\chi}-\Gamma_{\phi\overline{b\chi}}}{\Gamma}~\sim~\frac{4\lambda'\mu^2}{m_\phi\Gamma}{\rm Re}\left(\mathcal{I}\right){\rm Im}\left(\lambda\lambda''^* \right),
\label{eps}
\end{aligned}
\end{equation}
where the factor $\mathcal{I}$ involves the kinematics of the loop. The forms of $\epsilon$ and $\rm{Re}(\mathcal{I})$ are derived in Appendix \ref{A3}. 
Let us also parameterise the CP-violation in the decay widths as follows
\begin{equation}
\Gamma_{\phi b\chi}=\left(\frac{1+\epsilon}{2}\right)\Gamma_0, \qquad
\Gamma_{\phi \overline{b\chi}}=\left(\frac{1-\epsilon}{2}\right)\Gamma_0,
\label{Ga0}
\end{equation}
in terms of $\Gamma_0$, the tree-level partial width: $\Gamma_0\equiv\Gamma_{\phi \rightarrow b\chi}\big|_{\epsilon=0}=\Gamma_{\phi \rightarrow\overline{b\chi}}\big|_{\epsilon=0}$. Also, it should be noted that  $\Gamma_0=\Gamma_{\phi b\chi}+\Gamma_{\phi \overline{b\chi}}$~.

We can use the principle of detailed balance (see e.g.~\cite{Kolb:1979qa,Kolb:1990vq}) to rewrite the phase space functions in the form:
$f_\phi =f_2f_3=f_b^{\rm{eq}}f_\chi^{\rm{eq}}= e^{- \frac{E_b+E_\chi}{T}}~$.
Moreover, we can express the product of the out-of-equilibrium states in the following manner
$f_bf_{\chi}=\exp\left(-\frac{E_b}{T}-\frac{E_{\chi}}{T_{\chi}}+\Delta\right),$
where the factor $\Delta$ accounts for the chemical potentials
$ \Delta\equiv\frac{\mu_b}{T}+\frac{\mu_{\chi}}{T_{\chi}} $.
The phase space densities for the antiparticles can be written $f_{\overline{b}}f_{\overline{\chi}}=e^{-2\Delta}f_bf_{\chi}$ and
the ratio $f_bf_{\chi}/f^{\rm{eq}}_bf^{\rm{eq}}_{\chi}$ can be expressed
\begin{equation}
\frac{f_b f_\chi}{ f_b^{\rm{eq}}f_\chi^{\rm{eq}}}
=\frac{f_b^{\rm{eq}}\left(f_\chi^{\rm{eq}}\right)^{\alpha_\chi^{-1}}e^{\Delta}}{ f_b^{\rm{eq}}f_\chi^{\rm{eq}}}
= e^{-\frac{E_\chi(1-\alpha_\chi)}{T_\chi}}e^{\Delta},
\label{B2}
\end{equation}
where we have introduced the quantity $\alpha_i\equiv\frac{T_i}{T}$. In the case $\alpha_\chi=1$, the hidden sector containing $\chi$ and the visible sector are in equilibrium, and $\alpha_\chi=0$ corresponds to the temperature of the hidden sector being zero. Assuming that the change in the temperature difference between the two sectors is slow, we may reasonably take $\alpha$ to be constant in our calculations. 

Using  eq.~(\ref{Ga0}) \& (\ref{B2}) we re-express $\mathcal{D}_\chi$ in the following form
\begin{equation}
\begin{aligned}
\mathcal{D}_{\chi}=~\Lambda^{\phi }_{b\chi}\frac{(\mu\lambda)^2}{2}f_b^{\rm{eq}}f_\chi^{\rm{eq}}
\Bigg[\Big[
(1+\epsilon)F_\phi-(1-\epsilon)F_+
\Big]
&- 
\Big[(1+\epsilon) - (1-\epsilon) F_+
\Big]\frac{\Gamma_{\phi 23}}{\Gamma}  
\\
&\quad- \frac{1}{2}
\Big[ (1+\epsilon)^2 F_- -(1-\epsilon)^2  F_+\Big]
\frac{\Gamma_0}{\Gamma} \Bigg],
\end{aligned}
\end{equation}
where we have defined
\begin{equation}
\begin{aligned}
 F_\phi=\frac{f_\phi}{ f_b^{\rm{eq}}f_\chi^{\rm{eq}}},
 \qquad \qquad
 F_+=\frac{f_b f_\chi}{ f_b^{\rm{eq}}f_\chi^{\rm{eq}}},
 \qquad \qquad
F_-=\frac{f_{\overline{b}} f_{\overline{\chi}}}{ f_b^{\rm{eq}}f_\chi^{\rm{eq}}}.
\label{Fpm}
\end{aligned}
\end{equation}
Thus to leading order in the CP-violating parameter $\epsilon$
\begin{equation}
\begin{aligned}
\mathcal{D}_{\chi}\simeq&~\Lambda^{\phi }_{b\chi}\frac{(\mu\lambda)^2}{4}f_b^{\rm{eq}}f_\chi^{\rm{eq}}
\Bigg[
2(1+\epsilon)\frac{\Gamma_{\phi 23}}{\Gamma}  
\Big[F_\phi -1 \Big]
+
\frac{\Gamma_0}{\Gamma} 
\Big[F_\phi - F_+ +
 (1+2\epsilon)\left(F_\phi - F_- \right)
\Big]
\Bigg].
\label{CC}
\end{aligned}
\end{equation}
The equivalent expression for $\overline{\chi}$, denoted $\mathcal{D}_{\overline{\chi}}$, is given by eq.~(\ref{CC}) with $\epsilon\rightarrow-\epsilon$ and $F_\pm\rightarrow F_\mp$.

\subsection{Scattering contributions}

The contribution to the Boltzmann equation from the scattering terms, which we denote $\mathcal{S}_\chi$, is given by 
\begin{equation}
\begin{aligned}
\mathcal{S}_\chi=~&\Lambda^{23}_{b\chi}\left(|M|^2_{23\rightarrow b\chi}f_2f_3(1+ f_b)(1+ f_\chi)-|M|^2_{b\chi\rightarrow 23}f_bf_\chi(1+ f_2)(1+ f_3)\right)\\
&~+\Lambda^{\overline{b\chi}}_{b\chi}\left(|M|^2_{\overline{b\chi}\rightarrow b\chi}f_{\overline{b}}f_{\overline{\chi}}(1+ f_b)(1+ f_\chi)-|M|^2_{b\chi\rightarrow \overline{b\chi}}f_bf_\chi(1+ f_{\overline{b}})(1+ f_{\overline{\chi}})\right),
\end{aligned} 
\end{equation}
where we retain the statistical factors $(1+ f_i)$. 
Following \cite{Hook:2011tk}, this expression can be simplified using the finite density unitarity relationship \cite{Hook:2011tk, Weinberg:1979bt}
\begin{equation}
\begin{aligned}
&\Lambda_{b\chi}^{23}|M|^2_{b\chi\rightarrow 23}(1+ f_2)(1+ f_3)+
\Lambda_{b\chi}^{\overline{b\chi}}|M|^2_{b\chi\rightarrow \overline{b\chi}}(1+ f_{\overline{b}})(1+ f_{\overline{\chi}})\\
&\hspace{30mm}
=\Lambda_{b\chi}^{23}|M|^2_{23\rightarrow b\chi}(1+ f_2)(1+ f_3)+
\Lambda_{b\chi}^{\overline{b\chi}}|M|^2_{\overline{b\chi}\rightarrow b\chi}(1+ f_{\overline{b}})(1+ f_{\overline{\chi}}),
\end{aligned}
\end{equation}
to obtain
\begin{equation}
\begin{aligned}
\mathcal{S}_\chi=~&\Lambda^{23}_{b\chi}|M|^2_{23\rightarrow b\chi}
\Big[f_2f_3(1+ f_b)(1+ f_\chi)-f_bf_\chi(1+ f_{2})(1+ f_{3})\Big].
\end{aligned} 
\end{equation}
Then using the following relationships for systems in (sector-wise) thermal equilibrium \cite{Hook:2011tk,Kolb:1979qa}
\begin{equation}
\begin{aligned}
\frac{f_2f_3}{(1+ f_2)(1+ f_3)} &=e^{-(E_2+E_3)/T}, 
\hspace{1cm}
\frac{f_bf_\chi}{(1+ f_b)(1+ f_\chi)} &=e^{-(E_b+E_\chi)/T}e^{\Delta},
\end{aligned}
\end{equation}
this reduces further to
\begin{equation}
\begin{aligned}
\mathcal{S}_\chi=~&\Lambda^{23}_{b\chi}|M|^2_{23\rightarrow b\chi}f_2f_3(1+ f_b)(1+ f_\chi)
\Big[1-e^{(E_2+E_3-E_b-E_\chi)/T}~e^{\Delta}\Big].
\label{Sc1}
\end{aligned} 
\end{equation}
Thus, by energy conservation $(E_2+E_3)-(E_b+E_\chi)=0$, the scattering contribution vanishes at zeroth order in $\Delta$, and similarly for $\mathcal{S}_{\overline{\chi}}$.

\section{The asymmetric yield}
\label{S3}

To study the evolution of the asymmetry we examine the difference between the Boltzmann equations for $\chi$ and $\overline{\chi}$
\begin{equation}
\begin{aligned}
\dot{n}_{\chi-\overline{\chi}}+3H n_{\chi-\overline{\chi}} 
&\equiv
\left(\dot{n}_{\chi}+3H n_{\chi}\right)-\left(\dot{n}_{\overline{\chi}}+3H n_{\overline{\chi}}\right)\\
&=\left(\mathcal{D}_\chi-\mathcal{D}_{\overline{\chi}}\right)
+
 \left(\mathcal{S}_{\chi}-\mathcal{S}_{\overline{\chi}}\right).
 \label{E2}
 \end{aligned}
 \end{equation}
We shall derive an explicit expression for the asymmetry to zeroth order in $\Delta$, which contains the chemical potentials. Following this we discuss the washout terms which arise at $\OO(\Delta)$.

\subsection{The asymmetric yield}

First we note that since the scattering contributions vanish at zeroth order in $\Delta$, the Boltzmann equation for the asymmetry is
\begin{equation}
\begin{aligned}
\dot{n}_{\chi-\overline{\chi}}+3H n_{\chi-\overline{\chi}} 
=\left(\mathcal{D}_\chi-\mathcal{D}_{\overline{\chi}}\right)
+
\OO(\Delta).
 \label{qw}
 \end{aligned}
 \end{equation}
The difference between eq.~(\ref{CC}) and the equivalent expression for $\overline{\chi}$ gives
\begin{equation}
\begin{aligned}
\mathcal{D}_{\chi}-\mathcal{D}_{\overline{\chi}}=&~\epsilon\Lambda^{\phi }_{b\chi}(\mu\lambda)^2f_b^{\rm{eq}}f_\chi^{\rm{eq}}
\Bigg[
\frac{\Gamma_{\phi 23}}{\Gamma}  
\Big[F_\phi -1 \Big]
+
\frac{\Gamma_0}{\Gamma} 
\Big[F_\phi- \frac{F_-+F_+}{2} \Big]
\Bigg],
\label{DDb}
\end{aligned}
\end{equation}
where we have used that $\Gamma=\Gamma_{\phi 23}+\Gamma_{0}$.
Using the principle of detailed balance this can be re-written to zeroth order in $\Delta$  as follows
\begin{equation}
\begin{aligned}
\dot{n}_{\chi-\overline{\chi}}+3H n_{\chi-\overline{\chi}} 
=&~\epsilon(\mu\lambda)^2\Lambda^{\phi }_{b\chi}
\Bigg[\frac{\Gamma_{\phi 23}}{\Gamma}  
\Big[e^{-\frac{E_b+ E_\chi}{T_\phi}}-e^{-\frac{E_b+ E_\chi}{T}} \Big]
+\frac{\Gamma_0}{\Gamma} 
\Big[e^{-\frac{E_b+ E_\chi}{T_\phi}}- e^{-\frac{\alpha_\chi E_b+ E_\chi}{T_\chi}}
\Big]\Bigg].
\label{BE2}
\end{aligned}
\end{equation}
Performing the various integrals (see Appendix \ref{A4}) we obtain an expression for the asymmetric yield $Y_{\chi-\overline{\chi}}\equiv\frac{n_{\chi-\overline{\chi}}}{S}$ (where $S$ is the entropy density) to zeroth order in $\Delta$
\begin{equation}
Y_{\chi-\overline{\chi}}\simeq 
\frac{45\epsilon(\mu \lambda)^2m_\phi^2}{64\pi^4}
\left(\frac{45M_{\rm{Pl}}}{(1.66)\sqrt{g_*^\rho}g_*^S}\right)
\Bigg[
\left(\alpha_\phi^7 
- \frac{\Gamma_{\phi b\chi}}{\Gamma}\right)\left(\frac{m_\phi^2-m_\chi^2}{m_\phi^7}\right)
-\alpha_\chi^6
\frac{\Gamma_{0}}{\Gamma}\left(\frac{M^2-m_\chi^2}{M^7}\right) 
\Bigg].
\label{cor}
\end{equation}
where
\begin{equation}
M^2\equiv \alpha_\chi\left[m_\phi^2-m_\chi^2\left(1-\frac{1}{\alpha_\chi}\right)- m_b^2\left(1-\alpha_\chi\right)\right]\simeq m_\chi^2 +\alpha_\chi\left( m_\phi^2-m_\chi^2\right).
\label{MM}
\end{equation}
Further, if we define the deviation from equilibrium $\delta_i\equiv(1-\alpha_i)$, and expand to first order in $\delta_\chi$ and $\delta_\phi$, eq.~(\ref{cor}) reduces to
\begin{equation}
\begin{aligned}
Y_{\chi-\overline{\chi}}&\simeq 
\frac{45\epsilon(\mu \lambda)^2m_\phi^2}{128\pi^4}
\left(\frac{45M_{\rm{Pl}}}{(1.66)\sqrt{g_*^\rho}g_*^S}\right)
\Bigg[
7
\delta_\chi\frac{\Gamma_{0}}{\Gamma}\left(\frac{m_\phi^4-m_\chi^4}{m_\phi^9}\right) 
-14\delta_\phi\left(\frac{m_\phi^2-m_\chi^2}{m_\phi^7}\right)
\Bigg].
\label{YY}
\end{aligned}
\end{equation}
This is our main technical result. 
From inspection of eq.~(\ref{YY}), it can be seen how the temperature difference between the sectors, parameterised by $\delta_i$, affects the efficiency of the freeze-in mechanism. In Sect.~\ref{S4} we shall take certain limits for the $\delta_i$ corresponding to various cases of interest.

\subsection{Washout}
\label{SWO}

Thus far we have neglected any washout terms that arise from $\OO(\Delta)$ corrections. Specifically, if we examine the difference of the scattering terms $\mathcal{S}_\chi-\mathcal{S}_{\overline{\chi}}$, cf.~eq.~(\ref{Sc1}), at leading order in $\Delta$, it is seen that these give a negative `washout' contribution, which acts to remove any asymmetry
\begin{equation}
\begin{aligned}
\mathcal{S}_{\chi}-S_{\overline{\chi}}&=\Lambda^{23}_{b\chi}f_2f_3\Bigg(|M|^2_{23\rightarrow b\chi}(1+ f_b)(1+ f_\chi)
(1-e^{\Delta})
-|M|^2_{23\rightarrow \overline{b\chi}}(1+ f_{\overline{b}})(1+ f_{\overline{\chi}})
(1-e^{-\Delta})\Bigg).
\end{aligned} 
\end{equation}
Neglecting statistical factors, working to first order in $\Delta$ and zeroth order in $\epsilon$ the above expression can be related to the thermally averaged scattering cross section for $B_2B_3\rightarrow B_b\chi$
\begin{equation}
\begin{aligned}
\mathcal{S}_{\chi}-S_{\overline{\chi}}\simeq-2\Delta\Lambda^{23}_{b\chi}f_2f_3|M|^2_{b\chi\rightarrow 23}
= 
-2\Delta \langle\sigma v\rangle_{b\chi\rightarrow 23}  n_{b}^{\rm eq} n^{\rm eq}_{\chi}.
\label{SDel}
\end{aligned} 
\end{equation}
From inspection of eq.~(\ref{DDb}), the leading washout terms from $\mathcal{D}_{\chi}-\mathcal{D}_{\overline{\chi}}$ only arise at $\OO(\Delta^2)$.
The quantity $\Delta$ can be related to the asymmetric yield \cite{Kolb:1990vq}
\begin{equation}
\begin{aligned}
\Delta\equiv\frac{\mu_b}{T}+\frac{\mu_\chi}{T_\chi}\simeq\frac{Y_{\chi-\overline{\chi}}}{Y_{\gamma}},
\label{muD}
\end{aligned}
\end{equation}
 where $Y_\gamma\equiv\frac{n_\gamma}{S}\simeq0.14$. It follows from eq.~(\ref{SDel}) \& (\ref{muD}) that the Boltzmann equation for the total asymmetry to $\OO(\Delta)$, as given in eq.~(\ref{qw}), can be expressed as
 \begin{equation}
  \begin{aligned}
\frac{{\rm d}Y_{\chi-\overline{\chi}}}{{\rm d}x}
 &\simeq\frac{1}{xSH}
\Big(\mathcal{D}_\chi-\mathcal{D}_{\overline{\chi}}
-2 \langle\sigma v\rangle_{b\chi\rightarrow 23}  n_{b}^{\rm eq} n^{\rm eq}_{\chi}\frac{Y_{\chi-\overline{\chi}}}{Y_\gamma}\Big),
\label{DS}
 \end{aligned}
 \end{equation}
 where $S=\frac{2\pi^2g_*^ST^3}{45}$ is the entropy density, $H=\frac{1.66\sqrt{g_*^\rho}T^2}{M_{\rm{Pl}}}$ is the Hubble constant and $x\equiv\frac{m_\phi}{T}$.
 Since $ \langle\sigma v\rangle_{b\chi\rightarrow 23}\propto|M|_{ b\chi\rightarrow23}$ the washout processes are proportional to $ (\lambda^2\lambda'^2)Y_{\chi-\overline{\chi}}$. Therefore washout is significantly suppressed relative to the asymmetry generation and should be generally negligible in this class of models.
We can check this statement by solving  eq.~(\ref{DS}),  following \cite{Cui:2011ab}
 \begin{equation}
  \begin{aligned}
Y_{\chi-\overline{\chi}}
&\simeq
\int_0^x{\rm d}x'~\frac{{\rm d}Y_{\chi-\overline{\chi}}^{(0)}}{{\rm d} x'}
\exp\Big[-2\int_{x'}^{x}{\rm d}x''\frac{S(x'') \langle\sigma v\rangle_{b\chi\rightarrow 23}  Y_{b}^{\rm eq} Y^{\rm eq}_{\chi}}{x''H(x'')Y_\gamma}\Big].
\label{exp1}
 \end{aligned}
 \end{equation}
The argument of the exponential is the washout rate $\Gamma_{\rm WO}$ normalised to the Hubble rate
\begin{equation}
\frac{\Gamma_{\rm WO}}{H(x)}\equiv
\frac{S(x)Y_{b}^{\rm eq}Y^{\rm eq}_{\chi}}{xH(x)Y_\gamma}\langle\sigma v\rangle_{b\chi\rightarrow 23}.
\end{equation}
For $\Gamma_{\rm WO}/H<1$ the washout switches-off and the exponential factor in eq.~(\ref{exp1}) can be neglected. To obtain an estimate for the washout rate we take $\langle\sigma v\rangle_{b\chi\rightarrow 23}\sim [\lambda\lambda'\mu/m_\phi]^2/8\pi m_\chi^2$ and use $Y_{b}^{\rm eq}\simeq Y^{\rm eq}_{\chi}\simeq0.5/g_*^S\simeq7\times10^{-3}$ \cite{Kolb:1990vq} for complex scalars (valid in the regime $x\ll 3$) 
\begin{equation}
\begin{aligned}
\frac{\Gamma_{\rm WO}}{H(x)}
&\sim
\frac{\lambda^2\lambda'^2Y_{b}^{\rm eq} Y^{\rm eq}_{\chi}}{x^2}\left(\frac{\mu}{m_\phi}\right)^2
\left(\frac{M_{\rm{Pl}}}{m_\chi}\right)\\
&\sim10^{-16}
\left(\frac{1}{x}\right)^2
\left(\frac{\lambda}{10^{-7}}\right)^2
\left(\frac{\lambda'}{10^{-7}}\right)^2
\left(\frac{\mu}{1~{\rm GeV}}\right)^2
\left(\frac{10~{\rm GeV}}{m_\phi}\right)^2
\left(\frac{1~{\rm GeV}}{m_\chi}\right).
\label{GH}
\end{aligned}
\end{equation}
Dominant freeze-in production typically occurs around $x\sim2-5$ \cite{Hall} and, as typically $\Gamma_{\rm WO}/H\ll1$, the washout processes can be generally neglected in this class of models. Therefore the asymmetric yield to zeroth order in $\Delta$, as given in eq.~(\ref{YY}), provides a good approximation. The washout is negligible due to the fact that all processes which change $n_{\chi-\bar\chi}$ involve the feeble intersector couplings.


\section{Discussion}
\label{S4}

The asymmetric yield, eq.~(\ref{YY}), in the limit $m_\chi\ll m_\phi$ is of the form
\begin{equation}
\begin{aligned}
Y_{\chi-\overline{\chi}}&\simeq 
\frac{45\epsilon(\mu \lambda)^2}{128\pi^4 m_\phi^3}
\left(\frac{45M_{\rm{Pl}}}{(1.66)\sqrt{g_*^\rho}g_*^S}\right)
\Bigg[
7
\delta_\chi\frac{\Gamma_{0}}{\Gamma} -14\delta_\phi
\Bigg].
\label{YY2}
\end{aligned}
\end{equation}
In this section we examine the implications of this result. Firstly we shall confirm that if the mediator is brought into equilibrium, then the yield is significantly suppressed. Subsequently, we shall look to identify regimes in which asymmetries of similar magnitude to the observed baryon asymmetry can be generated. Finally, we consider a variant on the original formulation in which the DM hidden sector containing $\chi$ is in equilibrium with the visible sector and show that in this case the out-of-equilibrium mediator can still lead to an asymmetry being generated.

\subsection{Mediator in equilibrium}

Let us first examine the asymmetric yield for the original scenario of AFI \cite{Hall:2010jx}. Suppose that $\phi$ is in contact with the thermal bath due to an $\OO(1)$ coupling $\lambda'$ which dresses the interaction $\phi B_2B_3$, whereas the DM $\chi$ is thermally decoupled with $\lambda,\lambda''\lesssim10^{-7}$. This implies that the total width of $\phi$ is dominated by decays to the bath states $\Gamma\sim \lambda'^2\sim \OO(1)$. Moreover, the partial rate $\phi\rightarrow b\chi$ is parametrically
 \begin{equation}
\frac{\Gamma_{0}}{\Gamma}=\frac{\Gamma_{0}}{\Gamma_{0}+\Gamma_{\phi23}}\sim
\left(\frac{\lambda}{\lambda'}\right)^2\sim\lambda^2.
\end{equation}
As $\phi$ is in thermal contact with the visible sector, $T_\phi=T$. By inspection of eq.~(\ref{YY2}) with  $\delta_\phi=0$, it follows that the asymmetric yield is $\OO(\lambda^4)$ in the feeble coupling\footnote{Comparing with related analyses of leptogenesis using the Closed Time Path formalism e.g.~\cite{Garbrecht:2013iga}, one might expect that the asymmetry should actually vanish exactly in this case, rather than just at leading order. This distinction is unimportant for the question at hand, however we hope to return to this issue in future work.}  $\lambda\lesssim10^{-7}$, and thus highly suppressed (in agreement with \cite{Hook:2011tk}). Also, compared to the symmetric yield the asymmetry is significantly reduced in this scenario
\begin{equation}
\begin{aligned}
Y_{\chi-\overline{\chi}}\sim \epsilon\lambda^2 Y_{\chi+\overline{\chi}}.
\label{3.2}
\end{aligned}
\end{equation}
Moreover, $Y_{\chi-\overline{\chi}}$ is substantially smaller than the observed value of the asymmetry between baryons and anti-baryons $Y_{B-\overline{B}}\simeq 0.86\times10^{-10}$, being of order 
 \begin{equation}
\begin{aligned}
Y_{\chi-\overline{\chi}}&\sim 10^{-17}\left(\frac{\epsilon}{10^{-2}}\right)  
\left(\frac{\lambda}{10^{-7}}\right)^2\left(\frac{\mu}{\rm GeV}\right)^2  
\left(\frac{\Gamma_0/\Gamma}{10^{-14}}\right)
\left(\frac{\delta}{1}\right) \left(\frac{10~{\rm GeV}}{m_\phi}\right)^3~.
\label{Y1}
\end{aligned}
\end{equation}
We conclude that this minimal scenario can not produce sizeable asymmetries.

\subsection{Out-of-equilibrium mediator}

We next consider the scenario with $T_\phi,T_\chi\neq T$. As remarked earlier, it has been argued that the root cause of the suppressed asymmetry in the case $T_\phi=T$ can be traced to the fact that the mediator remains in thermal equilibrium \cite{Garbrecht:2013iga}. Henceforth we examine the case in which the mediator is out-of-equilibrium with the visible sector, which necessarily implies that the couplings $\lambda,\lambda',\lambda''\lesssim10^{-7}$. 
Thus we consider the asymmetric yield eq.~(\ref{YY2}) with $\delta_\phi\neq0$. 
Observe that for $\delta\chi$, $\delta\phi$, $\Gamma_0/\Gamma\sim1$ the asymmetry is suppressed relative to the symmetric yield only by the size of the CP-violation parameter
\begin{equation}
\begin{aligned}
Y_{\chi-\overline{\chi}}\sim\epsilon Y_{\chi+\overline{\chi}},
\label{3.3}
\end{aligned}
\end{equation}
and thus is enhanced compared to the previous case, cf.~eq.~(\ref{Y1}).
In this model all of the couplings involved in the freeze-in process are feeble, as a result the size of the CP-violation is generally smaller than in the original AFI scenario, comparing with eq.~(\ref{eps})
\begin{equation}
\begin{aligned}
\epsilon&
\sim
\frac{4\lambda'\mu^2}{m_\phi\Gamma}{\rm Re}\left(\mathcal{I}\right){\rm Im}\left(\lambda\lambda''^* \right)
~\sim~\frac{\lambda'~\rm{Im}(\lambda\lambda''{}^*)}{|\lambda|^2}
\lesssim 10^{-7}.
\label{eps2}
\end{aligned}
\end{equation}
Provided that washout is small (as we have argued previously), this can still potentially provide an asymmetry comparable to the observed asymmetry in baryon number.
Inspecting eq.~(\ref{YY2}), note that the contributions proportional to $\delta_\chi$ and $\delta_\phi$ oppose each other, and thus to generate a substantial particle asymmetry it is desirable for one of these terms to dominate the asymmetric yield. Let us start by examining the case that the first term gives the leading contribution to the asymmetric yield, i.e.~we consider the limit $\delta_\phi\sim0$. This scenario is much like that studied in the previous section, except in this scenario it is expected that $\Gamma_0/\Gamma\sim1$ and thus
 \begin{equation}
\begin{aligned}
Y_{\chi-\overline{\chi}}&\simeq 
10^{-10}\left(\frac{\epsilon}{10^{-8}}\right)  \left(\frac{\lambda}{10^{-7}}\right)^2
\left(\frac{\mu}{\rm GeV}\right)^2  
\left(\frac{\Gamma_0/\Gamma}{0.1}\right)
\left(\frac{\delta_\chi}{1}\right) \left(\frac{10~{\rm GeV}}{m_\phi}\right)^3~.
\label{Y2}
\end{aligned}
\end{equation}
We observe that in this case, even though the CP-violating coupling $\epsilon$ is small, sufficiently large asymmetries can in principle be generated. Moreover, provided that the only washout processes are due to these intersector interactions, the washout will be highly suppressed and this asymmetry can likely be retained. Thus in realistic models of AFI, it is important that this asymmetry is only frozen-in well after washout effects associated to electroweak symmetry breaking have frozen-out. This will likely be the case if the particle masses are all around the GeV scale.

\subsection{Freezing-in an asymmetry}
\label{S4.3}

Finally, we consider the other limit of eq.~(\ref{YY2}), in which the term proportional to $\delta_\phi$ dominates, as is the case if
\begin{equation}
\frac{1}{2}\left(\frac{\delta_\chi}{\delta_\phi}\right)\left(\frac{\Gamma_0}{\Gamma}\right)\ll1.
\end{equation} 
Note that this is realised if the mediator is in thermal equilibrium with the hidden sector, $\delta_\phi=\delta_\chi$. Comparing with eq.~(\ref{YY2}), in this scenario the asymmetry is reversed, leading to an excess in $n_{\overline{\chi}}$ over $n_\chi$ for $\epsilon>0$. However, the sign of $\epsilon$ is not fixed, it depends on the phase of $\lambda\lambda''{}^*$,  and one can always choose to redefine the CP parameter $\epsilon\rightarrow-\epsilon$ in eq.~(\ref{eps}). More importantly, the magnitude of the asymmetry is 
\begin{equation}
\begin{aligned}
\left|Y_{\chi-\overline{\chi}}\right|&\simeq 
\frac{315\delta_\phi |\epsilon|(\mu \lambda)^2}{64\pi^4 m_\phi^3}
\left(\frac{45M_{\rm{Pl}}}{(1.66)\sqrt{g_*^\rho}g_*^S}\right)~.
\label{YY3}
\end{aligned}
\end{equation}
In particular note that this term is $\OO(\lambda^2)$ and does not feature any factors of partial widths. An asymmetry is generated of a similar magnitude to eq.~(\ref{Y2}) 
\begin{equation}
\begin{aligned}
|Y_{\chi-\overline{\chi}}|&\simeq 10^{-10}\left(\frac{|\epsilon|}{10^{-8}}\right)  \left(\frac{\lambda}{2\times10^{-8}}\right)^2\left(\frac{\delta_\phi}{1}\right) 
\left(\frac{\mu}{\rm GeV}\right)^2 \left(\frac{10~{\rm GeV}}{m_\phi}\right)^3.
\label{YY4}
\end{aligned}
\end{equation}

Finally, a limit of particular interest is the case that $T_\chi =T$ and $T_\phi \neq T$. This implies that the hidden sector containing the DM $\chi$ and the visible sector are maintained at the same temperature (this may be because of additional interactions which do not alter the asymmetries or due to similar thermal evolutions after decoupling). As $\delta_\chi=0$ the asymmetric yield is given by eq.~(\ref{YY3}). Asymmetries in $\chi$ and $B_b$ can still be generated with $\chi$ in equilibrium with the bath, provided that the mediator is out-of-equilibrium $\delta_\phi\neq0$. In this scenario the DM is not frozen-in, but an asymmetry is. The interpretation is that the state $\phi$ freezes-in, but subsequently decays in a manner such that it generates a particle-antiparticle asymmetry. This is an interesting and, as we see in eq.~(\ref{YY4}), viable alternative to conventional AFI.

\section{Concluding remarks}
\label{S5}

AFI presents an alternate framework in which to understand the DM relic density and might allow an explanation of the cosmological coincidence $\Omega_{\rm{DM}}\approx5\Omega_B$. It has been previously suggested \cite{Hook:2011tk,Garbrecht:2013iga} that sizeable asymmetries can not be generated in AFI if the mediator is part of the visible sector thermal bath, thus here we have considered the scenario in which the mediator is out-of-equilibrium. From a careful analysis of the Boltzmann equations, we derived an expression for the asymmetric yield in this case, eq.~(\ref{YY}), and investigated how the difference in temperature of the mediator and the DM from the visible sector affects the magnitude of the asymmetry. Furthermore, in Sect.~\ref{SWO} we argued that if the washout only occurs due to intersector processes, then it is substantially suppressed and can typically be neglected.  In Sect.~\ref{S4} we showed that if the mediator is out-of-equilibrium with the thermal bath then the AFI mechanism can potentially give suitably large particle asymmetries, comparable to that observed in baryons $Y_{B-\overline{B}}\sim10^{-10}$. We also highlighted an interesting variant of the AFI paradigm in which the DM remains in thermal equilibrium with the bath, however an asymmetry is generated through the production (via freeze-in), and subsequent CP-violating decays, of an out-of-equilibrium state.

For clarity and to allow comparison with the earlier analyses of \cite{Hall:2010jx,Hook:2011tk}, we have restricted our attention to a simplified scalar model of AFI.  In order for the asymmetry to determine the relic density it is important that the symmetric component $Y_{\chi+\overline{\chi}}$, which is generally larger than the asymmetry by a factor of $1/\epsilon$, annihilates efficiently. Here we have only detailed the parts of the model relevant to the generation of particle asymmetries, cf.~eq.~(\ref{L}), however in a complete model there should be strong intrasector interactions which lead to the removal of the symmetric component. Replacing the scalar bath states with Standard Model particles should be straightforward exercise and should realise a simple viable model of AFI. The more ambitious challenge would be to construct a model in the context of a motivated framework of beyond the Standard Model physics (e.g.~supersymmetry). Phenomenological studies of specific realisations would also be of interest. 

In the models presented, for a sizeable asymmetry to be generated, whilst avoiding sector equilibrium, the couplings typically need to be of order $10^{-8\pm1}$. This is intriguing as although these couplings are small it is not infeasible that they might be probed by experiments, see \cite{Hall,FI} for relevant remarks. Further, supplementing the mediator sector (containing $\phi$) with additional states, which introduce new sources of CP-violation, could provide an interesting variant as it may be possible to enhance the size of $\epsilon$. Larger values of $\epsilon$ would allow for smaller intersector couplings whilst maintaining a suitable asymmetric yield. Additionally, a useful check of our conclusions could be made by using the Closed Time Path formalism, see e.g.~\cite{Garbrecht:2013iga}, rather than the Boltzmann analysis presented here. We hope to return to these issues in future publications.

\section*{Acknowledgements}

We are grateful to John March-Russell and Stephen West for early collaboration, and to Matthew McCullough and Michael Ramsey-Musolf for reading a draft. Further, we would like to thank Lawrence Hall, Anson Hook, Christopher McCabe, and Brian Shuve for useful discussions. 
 This work was initiated at the University Oxford, partially funded by an EPSRC studentship, and was completed at Santa Barbara KITP, supported in part by the National Science Foundation under Grant No.~NSF PHY11-25915.

\appendix

\section{Appendices}

\subsection{Construction of the Lagrangian}
\label{A1}

Following \cite{Hall:2010jx}, we outline the symmetry structure and field content which gives the Lagrangian studied in Sect.~\ref{S2}. Consider a model with complex scalar states $B_b$, $\Phi$ and $\chi$ and real scalars $\phi,B_2,B_3$. The states $B_b$ and  $\chi$ carry a conserved quantum number. Supposing $\phi$, $B_3$ and $\chi$ possess a parity symmetry (i.e $\chi\rightarrow-\chi$, etc.), we construct the following Lagrangian 
\begin{equation}
\begin{aligned}
\hat{\mathcal{L}}=\mu\Big(
\lambda \phi B_b \chi^* + \lambda' \phi B_2B_3
+\kappa_1 B_2\Phi^* B_b
+\kappa_2 B_3\Phi \chi^*
\Big)+{\rm h.c.}+\cdots~,
\end{aligned}
\end{equation}
where we omit terms which will not affect the Boltzmann equations.
To obtain a simplified setting which contains all of the necessary properties to demonstrate the mechanics of AFI we assume that  $\Phi$ is heavy and can be integrated out, to obtain the Lagrangian of Sect.~\ref{S2}
\begin{equation}
\begin{aligned}
\mathcal{L}=\mu\Big(
\lambda \phi B_b^* \chi + \lambda'\phi B_2B_3
\Big)
+\lambda'' B_2B_3 B_b^* \chi
+{\rm h.c.}+\cdots~.
\end{aligned}
\end{equation}
The coupling $\lambda'' =\kappa_1\kappa_2\mu^2/m_\Phi^2$ 
in terms of the fundamental couplings.
\\

\subsection{Relation for RIS subtraction}
\label{A2}

Here we prove a relation which we use in Sect.~\ref{S2}:
\begin{equation}
\Lambda^{ij}_{\alpha\beta}\frac{|M|^2_{ij\rightarrow \alpha\beta}  \Big|_{\rm{RIS}}}{|M|^2_{\phi\rightarrow \alpha\beta}}=\Lambda^{\phi}_{\alpha\beta}\frac{\Gamma_{\phi\overline{ij}}}{\Gamma}.
\label{AE4}
\end{equation}
This is valid for on-shell scatterings between pairs of states $ij$ 
with final states $\alpha\beta$ which are mediated by a single state $\phi$. The partial widths are of the form
\begin{equation}
\begin{aligned}
\Gamma_{\phi ij}&=\frac{1}{2m_\phi}\int{\rm{d}}\Pi_i{\rm{d}}\Pi_j(2\pi)^4\delta^{(4)}\left(p_\phi-p_i-p_j\right)|M|^2_{\phi\rightarrow ij}.\\\end{aligned}
\end{equation}
For on-shell scatterings the matrix element may be decomposed in the narrow width approximation as follows
\begin{equation}
\begin{aligned}
|M|^2_{ ij\rightarrow \alpha\beta} \Big|_{\rm{RIS}}&\simeq|M|^2_{ ij \rightarrow \phi}
\frac{\pi\delta(s-m_\phi^2)}{m_\phi\Gamma}|M|^2_{\phi\rightarrow \alpha\beta}.\\
\end{aligned}
\end{equation}
Thus we can write
\begin{equation*}
\begin{aligned}
\Lambda^{\phi}_{\alpha\beta}\frac{\Gamma_{\phi \overline{ij}}}{\Gamma}
&= \int \der \Pi_\phi \der \Pi_\alpha \der \Pi_\beta (2\pi)^4\delta^{(4)}\left(p_\phi-p_\alpha-p_\beta\right)\frac{\Gamma_{\phi\overline{ij}}}{\Gamma}\\
 &=
 \int \der \Pi_i \der \Pi_j \der \Pi_\alpha \der \Pi_\beta(2\pi)^4\delta^{(4)}\left(p_i+p_j-p_\alpha-p_\beta\right)|M|^2_{ij\rightarrow \phi}~\frac{\pi\delta(s-m_\phi^2)}{m_\phi\Gamma}\\
 &=
 \int \der \Pi_i \der \Pi_j \der \Pi_\alpha \der \Pi_\beta (2\pi)^4\int\der^4p_\phi|M|^2_{ij\rightarrow \phi}~\frac{\pi\delta(s-m_\phi^2)}{m_\phi\Gamma}\\
&\hspace{4.5cm}  \times \delta^{(4)}\left(p_\phi-p_\alpha-p_\beta\right)
 \delta^{(4)}\left(p_i+p_j-p_\phi\right),
\end{aligned}
\end{equation*}
and use the identity 
$\int{\rm{d}}\Pi_\phi=\int\frac{{\rm{d}}^4p_\phi}{(2\pi)^4}~2\pi\delta(s-m_\phi^2)$ to obtain
\begin{equation*}
\begin{aligned}
 \int \der \Pi_\phi \der \Pi_i \der \Pi_j \der \Pi_\alpha \der \Pi_{\beta}\frac{|M|^2_{ij\rightarrow \phi}}{2m_\phi\Gamma}  (2\pi)^8\delta^{(4)}\left(p_\phi-p_\alpha-p_\beta\right)
 \delta^{(4)}\left(p_i+p_j-p_\phi\right)
 =\Lambda^{ij}_{\alpha\beta}\frac{|M|^2_{ij\rightarrow \alpha\beta}  \Big|_{\rm{RIS}}}{|M|^2_{\phi\rightarrow \alpha\beta}}.
 \end{aligned}
\end{equation*}

\subsection{Magnitude of the CP-violation}
\label{A3}

In Sect.~\ref{S2} we stated the magnitude of the CP-violating parameter $\epsilon$ corresponding to the process of Fig.~\ref{Fig2}.
In this appendix we shall derive the expression for of $\epsilon$ given in eq.~(\ref{eps}) and, subsequently, calculate the form of $\rm{Re}(\mathcal{I}^*)$ in this model. We begin from the definition of the CP-violating parameter
\begin{equation}
\epsilon=~\frac{\Gamma_{\phi b\chi}-\Gamma_{\phi \overline{b\chi}}}{\Gamma}.
\label{A6}
\end{equation}
The leading contributions to the CP-violation arise from the diagrams shown in Fig.~\ref{Fig2}, thus
\begin{equation}
\begin{aligned}
\Gamma_{\phi b\chi}~\sim~\frac{1}{m_\phi}|-i\lambda\mu+(-i)^2\lambda'\mu\lambda''\mathcal{I}|^2
~\sim~ \frac{i\lambda'\mu^2}{m_\phi}\left(\lambda\lambda''^*\mathcal{I}^*-\lambda^*\lambda''\mathcal{I}\right)+\cdots~,
\end{aligned}
\end{equation}
where the ellipsis indicate terms which do no violate CP.  We can rewrite this as follows
\begin{equation}
\begin{aligned}
\Gamma_{\phi b\chi}~\sim~ \frac{2\lambda'\mu^2}{m_\phi}{\rm Im}\left(\lambda\lambda''^*\mathcal{I}^*\right)+\cdots~.
\end{aligned}
\end{equation}
A similar expression can be obtained for $\Gamma_{\phi \overline{b\chi}}$, but with $\lambda\rightarrow\lambda^*$ and $\lambda''\rightarrow\lambda''^*$. Substituting these into eq.~(\ref{A6}), after some algebra, we obtain the result quoted in eq.~(\ref{eps})
\begin{equation}
\epsilon\sim
\frac{4\lambda'\mu^2}{m_\phi\Gamma}{\rm Re}\left(\mathcal{I}\right){\rm Im}\left(\lambda\lambda''^* \right).
\end{equation}
If one of the couplings, say $\lambda$, dominates the width, then $\Gamma\sim \mu^2\lambda^2/m_\phi$ and the above form for $\epsilon$ can be further simplified, as used in eq.~(\ref{eps2}). It remains to determine the form of the factor ${\rm Re}\left(\mathcal{I}^*\right)$, which contains the loop kinematics due to the momenta of states $B_2$ and $B_3$
\begin{equation}
\mathcal{I}=\int\frac{{\rm} d^4 l}{(2\pi)^4}~
\prod_{j=2,3}~\frac{i}{\left(\sqrt{s}/2-l\right)^2-m_j+i\varepsilon}.
\end{equation}
From this, by standard cutting rules, it follows 
\begin{equation}
\begin{aligned}
2{\rm Im}(i\mathcal{I})&=i^2\int\frac{{\rm} d^4 l}{(2\pi)^4}
\prod_{j=2,3}\left(-2\pi i\delta\left(\left(s/2-l\right)^2-m_j^2\right)\right)\\
&=\int\frac{{\rm} d^4 q_2}{(2\pi)^4}\int\frac{{\rm} d^4 q_3}{(2\pi)^4}~(2\pi)^4
\delta^{(4)}\left(q_2+q_3-\sqrt{s}\right)4\pi^2\prod_{j=2,3}\delta^{(4)}\left(q_j-m_j\right)\\[3pt]
&=\int\frac{{\rm} d^3 q_2}{2E_2(2\pi)^3}\int\frac{{\rm} d^3 q_3}{2E_3(2\pi)^3}~(2\pi)^4
\delta^{(4)}\left(q_2+q_3-\sqrt{s}\right)\\[3pt]
&=\frac{1}{4\pi}\frac{\sqrt{s-(m_2+m_3)^2}\sqrt{s-(m_2-m_3)^2}}{2s}~\theta\left(s-(m_2+m_3)^2\right),
\end{aligned}
\end{equation}
involving the Heaviside $\theta$-function. For simplicity we can suppose $m_2=m_3$ (although no symmetry enforces this as the states have different $Z_2$ parities), in this case the result simplifies to the following form
\begin{equation}
{\rm Re}(\mathcal{I})=\frac{1}{16\pi}\sqrt{\frac{s-4m_2^2}{s}}~\theta\left(s-4m_2^2\right).
\end{equation}
An imaginary part of $i\mathcal{I}$ is only generated when the internal states are on-shell. The quantity ${\rm Re}(\mathcal{I})$ is the factor which appears in the parametric expression of $\epsilon$ in eq.~(\ref{eps}).

\subsection{Integrating the Boltzmann equation}
\label{A4}

In this appendix we present the steps taken to integrate the Boltzmann equation and obtain the form of the asymmetric yield given in eq.~(\ref{cor}). Our starting point is eq.~(\ref{BE2}), which we rewrite as follows
\begin{equation}
\begin{aligned}
\dot{n}_{\chi-\overline{\chi}}+3H n_{\chi-\overline{\chi}} 
=\epsilon(\mu\lambda)^2
\Bigg[\frac{\Gamma_{\phi 23}}{\Gamma}  
\Big[\mathcal{I}_{T_\phi}
-\mathcal{I}_{T} \Big]
+\frac{\Gamma_0}{\Gamma} 
\Big[\mathcal{I}_{T_\phi}- \mathcal{I}_{T_\chi}
\Big]\Bigg],
\label{BE3}
\end{aligned}
\end{equation}
with
\begin{equation}
\begin{aligned}
\mathcal{I}_{T_\phi} & =
\int \der \Pi_\phi  \der \Pi_b  \der \Pi_\chi  (2\pi)^4\delta^{(4)}(p_\phi-p_b-p_\chi)
~
e^{-\frac{E_b+E_\chi}{T_\phi}}\\
\mathcal{I}_{T_\chi} & =
\int \der \Pi_\phi  \der \Pi_b  \der \Pi_\chi  (2\pi)^4\delta^{(4)}(p_\phi-p_b-p_\chi)
~
e^{-\frac{\alpha E_b+E_\chi}{T_\chi}}
\label{gt}
\end{aligned}
\end{equation}
and $\mathcal{I}_T=\mathcal{I}_{T_\phi}\big|_{T_\phi=T}$~.
First, let us examine $\mathcal{I}_{T_\phi}$; we can use the identity 
\begin{equation}
\begin{aligned}
\int{\rm{d}}\Pi_\phi=\int\frac{{\rm{d}}^4p_\phi}{(2\pi)^4}~2\pi\delta(s-m_\phi^2)
\label{iden}
\end{aligned}
\end{equation}
and evaluate the integral over momentum to write this as
\begin{equation}
\begin{aligned}
\mathcal{I}_{T_\phi}\simeq
\pi\int\der\Pi_b\der\Pi_\chi \delta(s-m_\phi^2)~e^{-\frac{E_b+E_\chi}{T_\phi}}.
\end{aligned}
\end{equation}
To proceed we use the result of \cite{Edsjo:1997bg} and re-express the above as
\begin{equation}
\begin{aligned}
\mathcal{I}_{T_\phi}\simeq 
 \frac{T_\phi}{8\pi^3}
\int^\infty_{(m_b+m_\chi)^2}\der s~ \delta(s-m_\phi^2)P(s)K_1\left(\frac{\sqrt{s}}{T_\phi}\right).
\end{aligned}
\end{equation}
We have defined the function $P$ as follows
\begin{equation}
\begin{aligned}
 P(s)&=\frac{1}{2\sqrt{s}}\sqrt{s-(m_b+m_\chi)^2}\sqrt{s-(m_b-m_\chi)^2}\simeq \frac{s-m_\chi^2}{2\sqrt{s}},
\label{K1}
 \end{aligned}
\end{equation}
where $s=m_b^2+m_\chi^2+2E_bE_\chi -2|p_b||p_\chi|\cos\theta$  has its usual form.
Assuming $m_\phi>m_\chi+m_b$ we evaluate the integral to obtain 
\begin{equation}
\begin{aligned}
\mathcal{I}_{T_\phi}\simeq 
\frac{\alpha_\phi T}{16\pi^3}
\left(\frac{m_\phi^2-m_\chi^2}{m_\phi}\right)K_1\left(\frac{m_\phi}{T_\phi}\right).
\end{aligned}
\end{equation}
Turning to the second integral,  using eq.~(\ref{iden}) again, this can be expressed as
\begin{equation}
\begin{aligned}
\mathcal{I}_{T_\chi}\simeq
 \pi \int\der\Pi_b\der\Pi_\chi \delta(s-m_\phi^2)e^{-\frac{\alpha_\chi E_b+E_\chi}{T_\chi}}.
\end{aligned}
\end{equation}
Then, with an appropriate redefinition of variables, similar to before we can rewrite this
\begin{equation}
\begin{aligned}
\mathcal{I}_{T_\chi}\simeq 
 \frac{T_\chi}{8\pi^3}
\frac{1}{\alpha_\chi^2}\int^{\infty}_{(\alpha_\chi m_b+m_\chi)^2}\der s^{\prime}~\delta(s-m_\phi^2)P(s') K_1\left(\frac{\sqrt{s^\prime}}{T_\chi}\right),
\end{aligned}
\end{equation}
where we have defined a scaled version of $s$ such that we can apply directly the result of \cite{Edsjo:1997bg} 
\begin{equation}
\begin{aligned}
 s^{\prime}&=\alpha_\chi^2m_b^2+m_\chi^2+2\alpha_\chi E_bE_\chi -2|\alpha_\chi p_b||p_\chi|\cos\theta.
 \end{aligned}
\end{equation}
This is related to $s$ by 
$s^\prime=\alpha_\chi s+\alpha_\chi m_b^2\left(\alpha_\chi-1\right)+m_\chi^2\left(1-\alpha_\chi\right).$ Defining $M$ as in eq.~(\ref{MM})
\begin{equation}
M^2\equiv \alpha_\chi\left[m_\phi^2-m_\chi^2\left(1-\frac{1}{\alpha_\chi}\right)- m_b^2\left(1-\alpha_\chi\right)\right]\simeq m_\chi^2 +\alpha_\chi\left( m_\phi^2-m_\chi^2\right),
\end{equation}
the $\delta$-function can be written in terms of $s^\prime$ by making the replacement $\delta(s-m_\phi^2)=\alpha_\chi\delta\left(s^\prime-M^2\right)$.
Then integrating we obtain
\begin{equation}
\begin{aligned}
\mathcal{I}_{T_\chi}\simeq 
\frac{T}{16\pi^3}
\left(\frac{M^2-m_\chi^2}{M}\right) K_1\left(\frac{ M}{T_\chi}\right).
\end{aligned}
\end{equation}
Substituting into eq.~(\ref{BE3}) we can express the Boltzmann equation in the form
\begin{align}
\notag
\dot{n}_{\chi-\overline{\chi}}+3H n_{\chi-\overline{\chi}} 
\simeq &~
\frac{\epsilon(\mu \lambda)^2T}{16\pi^3}\left(\frac{m_\phi^2-m_\chi^2}{m_\phi}\right)\Bigg[
\alpha_\phi K_1\left(\frac{m_\phi}{T_\phi}\right)
- \frac{\Gamma_{\phi b\chi}}{\Gamma} K_1\left(\frac{m_\phi}{T}\right)
\\
&\hspace{35mm}
-\frac{\Gamma_{0}}{\Gamma}\left(\frac{m_\phi(M^2-m_\chi^2)}{M(m_\phi^2-m_\chi^2)}\right) K_1\left(\frac{ M}{T_\chi}\right)
\Bigg].
\notag
\end{align}
where we have used $\Gamma=\Gamma_{\phi b\chi}+\Gamma_0$. Finally, to obtain the yield we need to perform the integral with respect to time, which we  recast as an integral with respect to temperature using $\dot{T}=-HT$ (valid for $\frac{\partial g}{\partial T}\approx0$) 
\begin{equation}
\begin{aligned}
Y_{\chi-\overline{\chi}}&\simeq \int\der T~
\frac{\epsilon(\mu \lambda)^2}{16\pi^3SH}\left(\frac{m_\phi^2-m_\chi^2}{m_\phi}\right)\Bigg[
\alpha_\phi K_1\left(\frac{m_\phi}{T_\phi}\right)
- \frac{\Gamma_{\phi b\chi}}{\Gamma} K_1\left(\frac{m_\phi}{T}\right)
\\&\hspace{60mm}
-\frac{\Gamma_{0}}{\Gamma}\left(\frac{m_\phi(M^2-m_\chi^2)}{M(m_\phi^2-m_\chi^2)}\right) K_1\left(\frac{ M}{T_\chi}\right)
\Bigg].
\label{dT}
\end{aligned}
\end{equation}
Recall the entropy density is given by $S=\frac{2\pi^2g_*^ST^3}{45}$ and $H=\frac{1.66\sqrt{g_*^\rho}T^2}{M_{\rm{Pl}}}$.
 Expressing eq.~(\ref{dT}) as an integral with respect to the inverse scaled temperature $x\equiv\frac{m_\phi}{T}$ gives
\begin{align}
Y_{\chi-\overline{\chi}}\simeq ~&
\frac{\epsilon(\mu \lambda)^2}{16\pi^3}\left(\frac{m_\phi^2-m_\chi^2}{m_\phi^5}\right)
\left(\frac{45M_{\rm{Pl}}}{(1.66)2\pi^2\sqrt{g_*^\rho}g_*^S}\right)\\[3pt]
&\times
\int^{\infty}_0\der x~x^5
\Bigg[
\alpha_\phi K_1\left(\frac{x}{\alpha_\phi}\right)
- \frac{\Gamma_{\phi b\chi}}{\Gamma} K_1\left(x\right)
-\frac{\Gamma_{0}}{\Gamma}\left(\frac{m_\phi(M^2-m_\chi^2)}{M(m_\phi^2-m_\chi^2)}\right) K_1\left(\frac{x M}{\alpha_\chi m_\phi}\right)
\Bigg].
\label{dx}
\notag
\end{align}
We integrate with the approximation that the $\alpha_i$ can be taken as constant 
to obtain the yield 
\begin{equation}
Y_{\chi-\overline{\chi}}
\simeq 
\frac{45\epsilon(\mu \lambda)^2m_\phi^2}{64\pi^4}
\left(\frac{45M_{\rm{Pl}}}{(1.66)\sqrt{g_*^\rho}g_*^S}\right)
\Bigg[
\left(\alpha_\phi^7 
- \frac{\Gamma_{\phi b\chi}}{\Gamma}\right)\left(\frac{m_\phi^2-m_\chi^2}{m_\phi^7}\right)
-\alpha_\chi^6
\frac{\Gamma_{0}}{\Gamma}\left(\frac{M^2-m_\chi^2}{M^7}\right) 
\Bigg].
\notag
\end{equation}
%
%
%
%
%
%


\begin{thebibliography}{0}
%
\vspace{-3mm}
%
\bibitem{ADM}
  S.~Nussinov,
  Phys.\ Lett.\  B {\bf 165}, 55 (1985).
%
G.~B.~Gelmini, L.~J.~Hall, and M.~J.~Lin, 
Nucl. Phys. B281 (1987)
726.
%
  R.~S.~Chivukula, T.~P.~Walker,
  Nucl.\ Phys.\  {\bf B329 } (1990)  445.
%
%
  S.~M.~Barr, R.~S.~Chivukula and E.~Farhi,
  Phys.\ Lett.\  B {\bf 241}, 387 (1990).
%
  D.~B.~Kaplan,
  Phys.\ Rev.\ Lett.\  {\bf 68}, 741 (1992).
%
  D.~Hooper, J.~March-Russell and S.~M.~West,
  Phys.\ Lett.\  B {\bf 605}, 228 (2005)
  [hep-ph/0410114].
%
%
    D.~E.~Kaplan, M.~A.~Luty and K.~M.~Zurek,
  Phys.\ Rev.\  D {\bf 79}, 115016 (2009)
  [0901.4117].
%
   H.~An,  S.~L.~Chen, R.~N.~Mohapatra and Y.~Zhang,
  JHEP {\bf 1003}, 124 (2010)
  [0911.4463].
%
  D.~E.~Kaplan, G.~Z.~Krnjaic, K.~R.~Rehermann and C.~M.~Wells,
  JCAP {\bf 1005} (2010) 021
  [0909.0753].
%
%
  N.~Haba and S.~Matsumoto,
Prog.\ Theor.\ Phys.\  {\bf 125} (2011) 1311  [1008.2487].
%
  M.~R.~Buckley and L.~Randall,
  JHEP {\bf 1109} (2011) 009 [1009.0270].
%
  B.~Dutta, J.~Kumar,
  Phys.\ Lett.\  {\bf B699}, 364-367 (2011).
  [1012.1341].
  %
  M.~Blennow, B.~Dasgupta, E.~Fernandez-Martinez and N.~Rius,
  JHEP {\bf 1103} (2011) 014
  [1009.3159].
  %
  A.~Falkowski, J.~T.~Ruderman and T.~Volansky,
  JHEP {\bf 1105} (2011) 106 [1101.4936].
  %
  C.~Arina and N.~Sahu,
  Nucl.\ Phys.\  B {\bf 854}, 666 (2012)
  [1108.3967].
  %
  M.~T.~Frandsen, S.~Sarkar and K.~Schmidt-Hoberg,
Phys.\ Rev.\ D {\bf 84} (2011) 051703  [1103.4350].
  %
  R.~Kitano, H.~Murayama and M.~Ratz,
  Phys.\ Lett.\  B {\bf 669}, 145 (2008)
  [0807.4313].
     %
  D.~E.~Morrissey, K.~Sigurdson and S.~Tulin,
  Phys.\ Rev.\ Lett.\  {\bf 105} (2010) 211304
  [1008.2399].
%
  C.~Cheung and K.~M.~Zurek,
  Phys.\ Rev.\ D {\bf 84} (2011) 035007
  [1105.4612].
%
  J.~March-Russell and M.~McCullough,
JCAP {\bf 1203} (2012) 019  [1106.4319].
%
  J.~J.~Heckman and S.~-J.~Rey,
  JHEP {\bf 1106} (2011) 120
  [1102.5346].
  %
%
  T.~Lin, H.~-B.~Yu and K.~M.~Zurek,
  Phys.\ Rev.\ D {\bf 85} (2012) 063503
  [1111.0293].
%
%
  N.~Okada and O.~Seto,
  Phys.\ Rev.\ D {\bf 86} (2012) 063525
  [1205.2844].
  %
  C.~Arina, R.~N.~Mohapatra and N.~Sahu,
  Phys.\ Lett.\ B {\bf 720} (2013) 130
  [1211.0435].
%
%
  H.~Davoudiasl and R.~N.~Mohapatra,
  New J.\ Phys.\  {\bf 14} (2012) 095011
  [1203.1247].
%
%
  J.~Unwin,
  JHEP {\bf 1306} (2013) 090
  [1212.1425].
%
  L.~Pearce and A.~Kusenko,
  Phys.\ Rev.\ D {\bf 87} (2013) 12,  123531
  [1303.7294].
%
  K.~Petraki and R.~R.~Volkas,
  Int.\ J.\ Mod.\ Phys.\ A {\bf 28} (2013) 1330028
  [1305.4939].
  %
  K.~M.~Zurek,
Phys.\ Rept.\  {\bf 537} (2014) 91  [1308.0338].
%
  I.~-W.~Kim and K.~M.~Zurek,
  Phys.\ Rev.\ D {\bf 89} (2014) 035008
  [1310.2617].
%
  P.~Fileviez PŽrez and H.~H.~Patel,
  Phys.\ Lett.\ B {\bf 731} (2014) 232
  [1311.6472].
  %
  Y.~Zhao and K.~M.~Zurek,
  [1401.7664].
%
  E.~Hardy, R.~Lasenby and J.~Unwin,
  [1402.4500].
%
  K.~Petraki, L.~Pearce and A.~Kusenko,
  [1403.1077].

\vspace{-0.4mm}
\bibitem{Hall:2010jx}
  L.~J.~Hall, J.~March-Russell, S.~M.~West,
  [1010.0245].
  
\vspace{-0.4mm}
\bibitem{symC}
  M.~R.~Buckley,
  Phys.\ Rev.\  D {\bf 84} (2011) 043510
 [1104.1429].
  %
  J.~March-Russell, J.~Unwin and S.~M.~West,
  JHEP {\bf 1208} (2012) 029,
  [1203.4854].
  
  \vspace{-0.4mm}
\bibitem{Cui:2011ab}
  Y.~Cui, L.~Randall and B.~Shuve,
  JHEP {\bf 1204} (2012) 075
  [1112.2704].
  
  \vspace{-0.4mm}
\bibitem{Hall} 
 L.~J.~Hall, K.~Jedamzik, J.~March-Russell and S.~M.~West,
  JHEP {\bf 1003}, 080 (2010)
  [0911.1120].
%
 

 \vspace{-0.4mm}
\bibitem{FI}
  C.~Cheung,
  G.~Elor, L.~J.~Hall and P.~Kumar,
  JHEP {\bf 1103} (2011) 042
  [1010.0022];
 %
  JHEP {\bf 1103} (2011) 085
  [1010.0024].
  X.~Chu, T.~Hambye and M.~H.~G.~Tytgat,
  JCAP {\bf 1205} (2012) 034
  [1112.0493].
%
  M.~Blennow, E.~Fernandez-Martinez and B.~Zaldivar,
  [1309.7348].
%
  P.~S.~Bhupal Dev, A.~Mazumdar and S.~Qutub,
  Physics {\bf 2} (2014) 26
  [1311.5297].

\vspace{-0.4mm}
\bibitem{Hook:2011tk}
  A.~Hook,
  Phys.\ Rev.\ D {\bf 84} (2011) 055003
  [1105.3728].

\vspace{-0.4mm}
\bibitem{Garbrecht:2013iga}
  B.~Garbrecht and M.~J.~Ramsey-Musolf,
  Nucl.\ Phys.\ B {\bf 882} (2014) 145
  [1307.0524].

\vspace{-0.4mm}
\bibitem{Edsjo:1997bg}
  J.~Edsjo and P.~Gondolo,
  Phys.\ Rev.\ D {\bf 56} (1997) 1879
  [hep-ph/9704361].
  
  \vspace{-0.4mm}
\bibitem{Kolb:1979qa}
E.~W.~Kolb and S.~Wolfram,
  Nucl.\ Phys.\ B {\bf 172} (1980) 224
   [Erratum-ibid.\ B {\bf 195} (1982) 542].
  
\vspace{-0.4mm}
\bibitem{Kolb:1990vq}
  E.~W.~Kolb and M.~S.~Turner,
  Front.\ Phys.\  {\bf 69} (1990) 1.

\vspace{-0.4mm}
\bibitem{Weinberg:1979bt}
  S.~Weinberg,
  Phys.\ Rev.\ Lett.\  {\bf 42} (1979) 850.
  


  
\end{thebibliography}
\end{document}